\documentclass{article}
\usepackage{arxiv}
\usepackage[utf8]{inputenc}
\usepackage[T1]{fontenc}
\usepackage{hyperref}
\usepackage{url}
\usepackage{booktabs}
\usepackage{amsfonts}
\usepackage{nicefrac}
\usepackage{graphicx}
\usepackage{natbib}
\usepackage{tabularx}
\usepackage{longtable}
\usepackage{array}
\usepackage{caption}

\title{Governance by Design: Architecting Agentic AI for Organizational Learning and Scalable Autonomy}

\author{
  Nelly Dux\thanks{Nelly Dux is a doctoral researcher at ESSEC Business School in Cergy, France. Her research examines the governance and organizational implications of artificial intelligence, with a focus on how organizations design, implement, and scale agentic AI systems in knowledge-intensive settings. Contact: \texttt{dux@essec.edu}} \\
  ESSEC Business School\\
  \texttt{dux@essec.edu} \\
  \And
  Cristina Alaimo\thanks{Cristina Alaimo is Associate Professor of Information Systems, Data Analytics, and Operations at ESSEC Business School. Her research investigates how data, digital platforms, and AI-driven systems reshape markets and organizations. She is co-author of \emph{Data Rules: Reinventing the Market Economy} (MIT Press, 2024) and serves as Senior Editor at \emph{Organization Studies} and \emph{European Journal of Information Systems}. Contact: \texttt{alaimo@essec.edu}} \\
  ESSEC Business School\\
  \texttt{alaimo@essec.edu} \\
  \And
  Philippe Roussiere\thanks{Philippe Roussiere is Global Lead for Innovation and AI at Accenture Research. He has over 25 years of experience in research leadership across client-facing and thought leadership projects, with a focus on advancing the application of artificial intelligence across industries. Contact: \texttt{philippe.roussiere@accenture.com}} \\
  Accenture Research\\
  \texttt{philippe.roussiere@accenture.com} \\
  \And
  Abhishek Kumar Mishra\thanks{Abhishek Kumar Mishra is a Senior Principal at Accenture Research. He has over 17 years of experience in data science, artificial intelligence, and enterprise-scale AI transformation; his work currently focuses on designing and deploying multi-agent and retrieval-augmented systems for large-scale knowledge workflows, with applications across multiple industries. Contact: \texttt{a.bz.kumar.mishra@accenture.com}} \\
  Accenture Research\\
  \texttt{a.bz.kumar.mishra@accenture.com} \\
}

\date{}

\begin{document}

\maketitle

\begin{abstract}
Agentic AI systems---systems that can pursue goals through multi-step planning and tool-mediated action with limited direct supervision---are moving from experimental prototypes to enterprise deployments. This transition introduces tensions in implementation, scaling, and governance: organizations seek scalable autonomy for knowledge and coordination work, yet must preserve accountability, safety, cost control, and responsibility as systems initiate actions, access enterprise data, and evolve through iterative updates. Building on an in-depth qualitative case of a large IT services company's 2025 development and staged rollout of an agentic system integrated with enterprise tools, we show that governance is implemented through concrete architectural and working arrangements that determine what the system is allowed to do, which tools and data it can use, how memory is handled, and how performance improvements are introduced over time. We then distill seven lessons that explain how to build effective governance into agentic AI during operationalization and scaling.\footnote{The authors thank David Kimble, Vincenzo Palermo, Charlotte Raut, and Patrick Connolly for their contributions to this study.}
\end{abstract}


\section{Introduction}

AI is moving from a behind-the-scenes prediction engine to a front-stage organizational actor, shaping how work gets done, how decisions are prepared, and how coordination happens across teams and systems. In information systems and management research, this shift has been studied as part of the broader organizational consequences of AI as a transformative digital technology that reshapes routines, capabilities, and governance arrangements \citep{berente2021managing, floridi2018ai4people, mikalef2021artificial, vassilakopoulou2022information}. Yet as organizations move from generative AI tools to agentic AI systems that can act autonomously, managing this transition requires architectural and governance decisions that treat accountability and control as inseparable from the technical architecture maturity and organizational learning processes \citep{wang2024survey}.

In this study, agentic AI refers to AI systems that pursue complex goals over time with limited direct supervision by performing multi-step behavior that is not fully specified in advance. Practically, agentic systems differ from standard LLM chat tools along five dimensions relevant for organizations: autonomy (independent execution), tool use (calling enterprise services and APIs), orchestration (sequencing workflows and agents), memory (carrying state across steps or sessions), and delegated decision-making (choosing actions or routes within set boundaries and policies). This definition emphasizes that agentic AI is explicitly designed to (a) generate a plan, (b) invoke tools, (c) produce intermediate artifacts, (d) revise actions based on outcomes, and (e) store or retrieve information from memory structures \citep{wang2024survey}.

From a governance standpoint, these properties mean that AI systems are no longer limited to producing suggestions but can initiate actions, access enterprise tools, and influence outcomes across workflows. This change raises practical governance and design questions about who remains responsible for decisions and consequences when system behavior extends beyond isolated outputs and into coordinated organizational processes. For instance, risk management programs built for earlier AI systems (including model validation, ethics review, and post-hoc monitoring) are strained when an AI tool can initiate actions, produce compound errors across steps, and change behavior through updates or feedback loops \citep{raji2020closing, selbst2019fairness, dignum2019responsible}.

Regulatory frameworks also increasingly require organizations to ensure documentation, transparency, traceability, human oversight, and robustness, which become more complex in multi-component agentic systems \citep{euaiact2024, nist2023ai}. Prior work provides important building blocks for integrating governance in agentic systems: typologies of LLM-based agents and multi-agent systems \citep{schick2023toolformer, wang2024survey}; technical patterns such as retrieval-augmented generation (RAG), tool calling, and agent frameworks; and organizational discussions of responsible AI and human--AI collaboration \citep{ananny2018seeing, berente2021managing, raisch2021artificial}.

Yet the scholarly and practical landscape remains fragmented on a central question: how governance is designed, enacted, and updated for agentic systems inside mature organizations, where systems must operate across heterogeneous tools, established data infrastructures, and organizational routines that do not move at the same pace as AI stacks \citep{alaimo2024data}. Therefore, we ask: \emph{How do organizations design, evolve, and scale governance for agentic AI systems as they acquire more operational autonomy in enterprise workflows.}

Rather than proposing a definitive governance framework, we highlight the nexus between system development and learning maturity and examine the central challenge that organizations face when implementing and scaling agentic AI: how to increase system autonomy while preserving accountability, traceability, and control. We show that governance in agentic AI differs from earlier technological and AI governance approaches because it should be progressively embedded into architectural design decisions concerning tool access, data boundaries, memory configuration, and human oversight checkpoints, rather than implemented as an external compliance layer.

Building on this insight, we introduce a learning maturity model showing that governance is not a static enactment of control but evolves alongside both the system's learning capabilities and the organization's readiness: as agents transition from push learning to explicit and implicit feedback learning, their growing capacity for adaptation requires increasingly sophisticated data governance, memory design, oversight mechanisms, and corresponding changes in user practices, implementation routines, and team configurations.

The remainder of the paper proceeds as follows. First, we describe our qualitative case study of a large IT services organization (hereafter, ``the organization'' or ``the company'') and outline the empirical setting, data sources, and analytical approach. We then present the findings in two steps. Step one reconstructs the staged evolution of the organization's agentic architecture, from bounded assistants to integrated workflows and agentic orchestration, showing how governance was progressively embedded into system design. Step two introduces the learning maturity model, which conceptualizes how different levels of memory and feedback integration reshape governance requirements over time.

Building on this foundation, we analyze four governance challenges that emerged during scaling: missing memory, tool and data fragmentation, learning without loss of traceability, and conditional autonomy at the workflow--agent boundary; we then explain how the company addressed each challenge through governance-by-design interventions. We conclude by synthesizing how memory configuration, data boundaries, and staged autonomy collectively turn agentic architecture into an organizational learning journey under accountable human oversight.

\section{Methods and Empirical Context}

\subsection{Research Setting}

This paper draws on a 2025 qualitative case study conducted at a large IT services organization. The company represents a theoretically informative setting because it combines high knowledge intensity with an experimentation-oriented culture in which speed, quality, and responsibility are core operating principles. The company operates as a vertically integrated organization, with research, engineering, design, and delivery capabilities available for project staffing---capabilities that enable rapid iteration from conceptual vision to deployable asset. This intrapreneurial mindset (moving from idea to outcome through iterative builds) creates strong internal momentum to adopt and scale emerging technologies such as agentic AI. Moreover, the company benefits from democratized access to technology and substantial corporate investment, allowing the team to leverage enterprise-grade infrastructure and AI tooling.

The company established an ``agentic lab'' around 2022, prior to the widespread availability of commercial large language models, as a dedicated unit to explore how purpose-built AI agents could augment internal research processes and enable new forms of productivity. While the lab develops internal technological assets, it is also frequently involved in client-facing engagements, which can interrupt design-and-build workflows, and introduce competing priorities between experimentation and delivery. This combination of autonomy, access to resources, focus on different use cases, and governance oversight makes this company and its agentic lab an analytically rich site for examining how agentic AI capabilities are piloted, stabilized, and translated into repeatable routines within a mature organization.

\subsection{Research Design and Data Sources}

We conducted an in-depth qualitative case study of both the company's development and staged implementation of an agentic AI system integrated with existing workflows and enterprise data sources. Data were collected through (a) observation of system build and scale-up activities, (b) interviews with key stakeholders involved in architectural design and organizational rollout, and (c) archival materials capturing system interfaces, architectural diagrams, and governance-relevant artifacts (e.g., plans, logging practices, and onboarding materials). The analysis focused on tracing how technical choices and organizational arrangements co-evolved as autonomy increased and as governance tension became salient.

Primary data comprised 17 semi-structured interviews with 12 employees across three waves: a pre-rollout wave (Wave 1, April--May 2025; $n = 9$), a post-rollout wave (Wave 2, July 2025; $n = 5$), and a follow-up wave (Wave 3, November 2025; $n = 3$), supplemented by one written interview conducted via email exchange. (See Table~\ref{tab:datasources} for an overview of the data sources and Table~\ref{tab:interviews} for an overview of interviews.) Additionally, three observations were conducted---two tool rollout training sessions (in-person and online, June 2025) and a joint research event (18 June 2025)---alongside eight update meetings with the core research team. Secondary data included 20 internal and public organizational documents (presentations, reports, RAI frameworks, press releases, and blogs), two industry articles, and one internal survey.

\begin{table}[htbp]
\centering
\caption{Overview of primary and secondary data sources}
\label{tab:datasources}
\begin{tabular}{lc}
\toprule
\textbf{Data sources} & \textbf{Amount} \\
\midrule
\textit{Primary data sources} & \\
Interviews & 18 \\
\quad -- Wave 1: pre-rollout (April--May 2025) & 9 \\
\quad -- Wave 2: post-rollout (July 2025) & 5 \\
\quad -- Wave 3: follow-ups (November 2025) & 3 \\
Written interview (email exchange) & 1 \\
Update meetings & 8 \\
Observations & 3 \\
\quad -- Tool rollout training sessions (in-person and online) & 2 \\
\quad -- ESSEC--organization research event (6/18/25) & 1 \\
\midrule
\textit{Secondary data sources} & \\
Internal documents (presentations, reports, RAI frameworks) & 7 \\
Public company documents (press releases, reports, blogs) & 10 \\
Industry and news articles & 2 \\
Internal surveys & 1 \\
\bottomrule
\end{tabular}
\end{table}

\begin{table}[htbp]
\centering
\caption{Overview of interviews}
\label{tab:interviews}
\small
\begin{tabular}{lll}
\toprule
\textbf{Interviewee} & \textbf{Date} & \textbf{Duration} \\
\midrule
Interviewee 1: Responsible AI Lead & 4/10; 5/19; 11/14/25 & 70 min; 30 min; 50 min \\
Interviewee 2: Research Executive & 4/11/25 & 26 min \\
Interviewee 3: Talent Coordinator & 4/16; 11/19/25 & 37 min; 25 min \\
Interviewee 4: Front-end Designer & 5/9/25 & 71 min \\
Interviewee 5: Industry Researcher (Resources) & 5/14/25 & 62 min \\
Interviewee 6: Industry Researcher (Travel) & 5/26/25 & 61 min \\
Interviewee 7: Senior Editor & 5/29; 7/1/25 & 61 min; 55 min \\
Interviewee 8: Editor/Narrative Designer & 5/30/25 & 75 min \\
Interviewee 9: Editorial Researcher & 7/11/25 & 55 min \\
Interviewee 10: Developer (back-end) & 7/15/25 & 45 min \\
Interviewee 11: Developer (back-end) & 7/17; 7/21; 11/5/25 & 60 min \\
Interviewee 12: Designer (front-end and UX) & 7/21/25 & 52 min \\
\bottomrule
\end{tabular}
\end{table}

\subsection{Analytical Approach}

We used an iterative thematic analysis approach to identify governance mechanisms and recurring tensions as the system evolved from bounded assistants toward a more agentic architecture. We first constructed a temporal map of key events and design decisions, organizing the empirical material along a timeline that distinguished three phases aligned with the system's staged rollout: a pre-rollout phase (March--May 2025), in which we captured baseline conditions and conducted ``wave 1'' interviews, as beta testing with $\sim$25 users was being established; a rollout and adoption phase (June--July 2025), coinciding with the first official deployment to company staff on June 4, 2025, during which we conducted field observations of training sessions and wave 2 interviews to document immediate user responses and emerging governance challenges; and a post-adoption and scaling phase (October 2025--February 2026), in which wave 3 follow-up interviews traced how architectural choices were revised, new governance mechanisms introduced (e.g., trust scoring, AI-assisted beta testing), and plans for broader organizational adoption were developed.

It is important to note that the underlying technological platform, initially a set of standalone AI assistants and internal tooling, had been in development within the company's Agentic Lab since approximately 2022, well before commercial large language models became widely available. The agentic capabilities described in our findings (orchestration, multi-agent coordination, governed data layers) emerged later, through the staged transformation that began in early 2025. Our study therefore captured the specific window in which this platform transitioned from bounded assistants to agentic execution, and in which governance tensions became salient.

Within each phase, we traced how architectural choices, governance arrangements, and user practices co-evolved. Themes were developed through repeated engagement with field material and refined against governance constructs from responsible AI and lifecycle governance literatures \citep{floridi2018ai4people, rahwan2018society}.

\section{Findings}

Our findings show that the main implementation challenge concerned the changing governance requirements as the agentic system gradually gained the capacity to plan, invoke tools, coordinate subtasks, and potentially retain state across sessions. The initial governance approach was mainly focused on solving the coordination bottlenecks arising from a fragmented tool environment, described internally as a ``spaghetti configuration'' (interviewee 1), where multiple assistants and workflows operated without integration. As autonomous capabilities expanded, however, the core governance issue shifted from increasing functionality to programming control, which, in agentic systems, translates into how to structure decision rights, memory configuration, and data access so that the system remains accountable, efficient, and operationally stable. Governance, therefore, was progressively embedded into architectural design and working routines that defined when and how the system could act.

We structure our findings in three parts. First, we reconstruct the organization's step-by-step evolution from isolated LLM-based tools toward a layered agentic AI architecture that integrates human oversight checkpoints, orchestration logic and reviewer agents, governed data layers, and differentiated workflow versus agent execution. Second, we introduce the learning maturity model developed inductively from the case, conceptualizing three levels of learning (1.~push learning, 2.~explicit feedback learning, and 3.~implicit feedback learning), while linking each level to specific configurations of memory, data governance, and instrumentation. Third, we synthesize the governance tensions through a challenge--solution lens, identifying four governance challenges that emerged during scaling and detailing how the organization addressed each with concrete design interventions aligned with the maturity model.

Together, our three-part findings demonstrate how agentic AI implementation substantially differs from classic system implementation \citep{orlikowski2007sociomaterial}, as the autonomous capabilities of agentic systems confer on the process the characteristics of an organizational learning journey that unfolds through iterative governance-by-design decisions. What makes this governance approach distinctive is that it does not operate as a static compliance layer applied to a finished system, but as an evolving sociotechnical arrangement that shapes both the system architecture and the organizational context in which it is implemented. This dynamic becomes visible as the system moves from operating on fixed knowledge sources toward incorporating feedback from user interaction and, eventually, learning from broader patterns of use.

Across this progression, governance evolves alongside both the technical system and the organization implementing it. As agents acquire stronger learning capabilities, organizations have to simultaneously adapt their data governance practices, architectural controls, user training, implementation routines, and team coordination mechanisms. Agentic AI governance thus emerges as a sociotechnical process in which autonomy expands only as the system's data, memory, and oversight foundations, as well as the organization's readiness to manage them, are sufficiently developed.

\subsection{Part A: The Three-Phase Evolution of Agentic AI}

The company did not set out to build AI agents merely in response to a new technological novelty. Instead, adoption followed an ``ROI-first logic'', one that treated agentic systems as an investment expected to improve research production measurably. Here, the main productivity constraint was perceived to be technological fragmentation as teams worked across and with disconnected tools, data sources, and point solutions, forcing analysts to move information between systems, reconcile inconsistencies, and assemble deliverables manually. One interviewee described the setup as a ``spaghetti configuration'' (interviewee 1), where context and data were spread across systems, making reuse difficult and consistency hard to maintain.

Analysts and editors spent time moving content between applications, repeating searches across databases, and reassembling drafts by hand. Without an integrated organizing platform, quality control and source compliance depended heavily on individual routines. As one researcher noted, ``\emph{everything was a little fragmented\ldots\ specific to teams' use. Each team used their own tools for their purpose}'' (interviewee 5), which sometimes contributed to duplication and misalignment. For example, editors recounted having to verify that citations from one system matched the figures generated in another.

The organization's leadership realized that simply adding more tools or rules wouldn't solve the ``spaghetti'' problem. Instead, they needed an integrated approach, one that could automate the handoffs and ensure continuity of context, so that each step of research didn't start from scratch. The solution would require rethinking how tasks flow through systems, as well as how data and context could be shared more seamlessly (and safely) across them. This goal oriented the team toward agentic AI as a means to reduce individual labor and manual rework, shorten end-to-end cycle times, increase throughput for recurring tasks, and tighten quality and risk control through more traceable workflows. As one interviewee noted: ``\emph{We had so many tools that it was difficult to put everything together\ldots\ This system design becomes very close to organization design, because it needs to implement decisions}'' (interviewee 10).

To solve the coordination bottleneck, the company embarked on a stage-based transformation in 2025. This meant proceeding through gradual transformation, with each phase introducing more AI initiatives, but within carefully bounded limits.

\paragraph{Stage 1: Bounded assistants.} The company's first step was to deploy generative AI assistants that were integrated into existing workflows. These were ``bounded'' assistants in the sense that they had narrow tasks (like drafting summaries or extracting insights) and operated under human guidance and review. For example, analysts could use an LLM-based helper to draft a section of a report. This boosted productivity in drafting and synthesis. However, the output of these assistants still underwent careful checking: the productivity value of the AI's help was thus directly tied to the team's time spent verifying its outputs and backing them up with sources.

To solve the issue, the organization learned to implement verification routines, including by developing standardized prompt templates that asked the AI to cite its sources and by instituting a norm that a human editor should review AI-generated text against the source material. In essence, stage 1 revealed that speed gains from AI required parallel improvements in process and governance rules, whereby the team formalized how to cross-check and fact-check the AI's contributions and standardized prompts. This stage established basic co-working norms between humans and AI, but it didn't yet solve the ``fragmentation'' problem---the assistants were still isolated in each tool.

\paragraph{Stage 2: Wrapping legacy tools as services.} The next step was oriented to address the ``integration'' issue. The company had many proven legacy utilities (for data analysis, visualization, etc.) that were valuable but isolated. The team began ``wrapping'' the functions of these legacy tools into AI callable services, essentially turning them into modular components that an agent-orchestrator could invoke via API, integrating their functionality within the agentic system. The strategic mindset was, however, to ``\emph{use automation only where it solves a concrete delivery problem}'' (interviewee 12). For instance, instead of users manually copying outputs into another artifact, the system would use API calls in the background to retrieve what it needed from other tools. Control would then pass back to a main orchestrator that sequenced steps across sources and components, reducing manual stitching. Next, the team built a prototype agent that could chain tasks and execute an end-to-end deliverable, such as ``generate a PPT slide deck with updated market figures,'' by invoking several services in the correct order.

These workflows remained largely deterministic and served as a deliberate starting point, motivated by a strategic cost decision. Workflow designs that require agent planning, control, and multi-step execution are, in fact, more expensive to build and run, and they are not always necessary to produce value. The company therefore opted to start with one-directional decision workflows to establish core controls and integration with minimal complexity and resource spending. This stage achieved integration across tools, since tasks that previously required several handoffs could now be executed in a single run. Only after that workflow baseline was stable did they consider adding complexity by identifying which workflows genuinely required deeper reasoning and adaptive planning.

\paragraph{Stage 3: Agentic orchestration.} By mid-2025, the company moved to agentic execution, in which some workflow systems not only followed predefined sequences but also planned and made choices. In this phase, the organization introduced a central orchestrator agent capable of interpreting a user's request, breaking it into subtasks, and coordinating multiple specialized sub-agents to deliver a result. For instance, given a complex query like ``Analyze the trend of renewable energy investments and produce a summary with charts and citations,'' the orchestrator might create a plan: (a) query internal investment databases, (b) summarize findings with an LLM, (c) generate charts with the plotting service, (d) have a reviewer agent check facts, and (e) compile the report.

This stage introduced partial autonomy in executing the workflow: the coordinating agent decides which tools (data agents, writing agents, etc.) to call and in what order. Importantly, the autonomy was bounded by design, as the agent could only call approved tools and sources, and it had to follow certain templates (e.g., always produce a citation for a fact, always run the ``hallucination-check'' agent on text). In Stage 3, the system executed a multi-step, data-driven workflow on the user's behalf by combining deterministic workflow logic with agentic execution. Agentic behavior became salient here because the system did act with a degree of autonomy. It sequenced actions across tools and data sources, allocated subtasks to specialist agents, and coordinated their outputs through an orchestrator that functioned like a project manager.

Each stage thus functioned as a pilot with explicit learning goals, making experimentation structured practice. In practical terms, the approach helped clarify which tasks are in scope, who the test users are, what success metrics look like, what failure modes to watch for, and how feedback will be captured and acted on. This staging created concrete learning loops, such as how users respond to AI drafts, where verification breaks down, which steps introduce errors when automation is added, and which controls are necessary before scaling. By building in stages, the company developed the technology and cultivated the organization's readiness at the same time.

The gradual rollout also created time to develop policies, onboarding materials, and training routines that matched the system's evolving capabilities. When stage 3 was piloted, users were already socialized to specify tasks clearly and validate outputs through the prompt-and-verify habits developed in stages 1 and 2---including earlier exposure to AI assistants like Copilot or ChatGPT Enterprise. The result is an iterative co-evolution dynamic in which tool capability and user practice advance together, with pilot cycles serving as the mechanism for adaptation and controlled scaling.

We summarize the organization's architectural agentic AI design below.

\bigskip
\noindent\fbox{\parbox{\textwidth}{
\textbf{State-of-the-art agentic architecture at a large IT services organization}

\medskip
By stage 3, the company's AI system had evolved into a three-layer architecture with a setup explicitly designed to embed governance into the technology. The layers consisted of a UI layer (user interface), an agent orchestration layer, and a data layer, with control points connecting them. Each layer was a technical component with embedded governance mechanisms (Figure~\ref{fig:architecture}).

\begin{center}
\includegraphics[width=\textwidth]{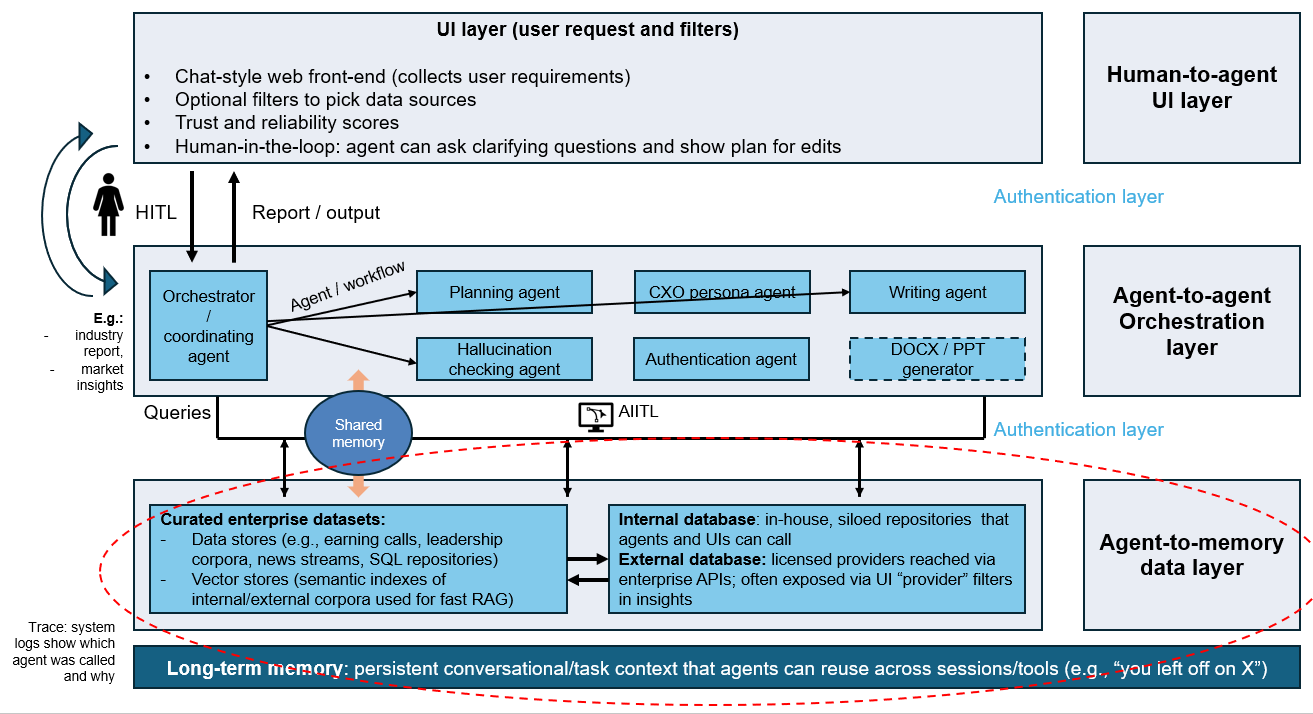}
\captionof{figure}{Agentic system architecture at the company}
\label{fig:architecture}
\end{center}

\paragraph{User interface layer---human oversight.} The company built a chat-style front-end where researchers interact with the system. When a user types a request, the system generates a proposed plan (essentially the AI's intended sequence of steps) and shows it to the user for approval. The researcher can edit the plan or adjust filters (e.g., narrowing the data sources to be searched) before execution. By making planning explicit and requiring user sign-off, the organization turned the UI into a governance checkpoint as it implements the principle of human-in-the-loop approval. It also educates users about the system's capabilities and limits, fostering a shared understanding of what the agent will do.

\paragraph{Agent orchestration layer---automated checks and modularity.} After a user approves the plan, execution shifts to an orchestration layer, where a coordinating agent delegates work to specialist sub agents, and governance is enforced through rules that determine which agent or tool handles a given request and under what conditions, so data access rules are implemented directly in the system's tool selection and source constraints. Quality assurance is handled by independent reviewer agents (``agents-in-the-loop'') that check outputs against retrieved sources, trigger rework when claims are unsupported, and run tiered review loops resembling an editorial process---thereby enabling agent-in-the-loop validation without always returning to a human. Finally, every action is logged with timestamps and references, creating an audit trail that supports traceability and accountability if issues emerge later.

\paragraph{Data layer---data scope, access and quality.} At the bottom lies the data layer, which determines what information agents can draw on. The organization's data layer comprises governed enterprise datasets and vetted external sources, organized into two types of storage: traditional data stores and vector stores. The data stores include internal repositories (research reports, survey data, etc.) and external content licensed for use (e.g., market research databases), all accessed through enterprise-approved APIs. On top of these, the organization built vector indices---essentially semantic search indexes---to allow agents to do retrieval-augmented generation (RAG) efficiently. The critical governance element here is that the AI's knowledge is limited to a controlled database. The system can only retrieve from these curated sources, and even within them, additional filters can apply.
}}
\subsection{Part B: Learning Maturity of AI Agents}

The IT services organization conceptualized its approach to developing agentic AI in terms of learning maturity levels. These levels describe how the agentic system incorporates learning from data or interaction with the environment, and each level corresponds to distinct memory needs.

\paragraph{Maturity level 1: push learning.} Level 1 describes the company's 2025 baseline, where the agent does not learn from use. Instead, improvements happen only when designers update curated knowledge sources, rules, and templates. Memory stays largely static and read-only at runtime, relying on fixed corpora and configuration, plus limited session context, so that the agent can retrieve information but does not write durable feedback or session history back into the system. Governance is therefore relatively straightforward because the agent's behavior does not change over time, unless humans deliberately change the underlying resources and protocols.

\paragraph{Maturity level 2: explicit feedback learning.} At this stage, users give explicit, structured feedback on specific outputs, and the system stores it in a feedback log linked to the output and its context. Over time, the agent can draw on this feedback memory, along with the static knowledge base, to adjust its responses to similar tasks. This requires traceability of structured feedback across the interface and orchestration layers, so that feedback can be attached to specific interaction or workflow steps. Governance also becomes more complex because user input now changes what the system remembers and potentially how it behaves toward others. Roles, attribution, and review mechanisms are thus needed to prevent low-quality behavioral adaptations from propagating.

\paragraph{Maturity level 3: implicit feedback learning.} Level 3 is the most ambitious stage, in which the system learns from implicit usage signals---such as edits, clicks, and ignored suggestions---rather than from explicit feedback. To do this, it would need to log end-to-end run histories across users and infer what should change, moving toward reinforcement learning-style adaptation. The organization described this as aspirational and currently not realistically reachable (2025--2026), given both the technical burden (data volume, learning methods) and the governance burden. Here, governance becomes the central bottleneck because implicit learning makes it harder to audit what the system has inferred, detect bias, and maintain accountability---all of which require strong logging, behavior versioning, and formal approval controls for self-learning changes.

We summarize the learning and maturity model of agents in Table~\ref{tab:maturity}.

\begin{table}[htbp]
\centering
\caption{Maturity level and learning type of agents}
\label{tab:maturity}
\small
\begin{tabularx}{\textwidth}{>{\raggedright\arraybackslash}p{2cm}>{\raggedright\arraybackslash}X>{\raggedright\arraybackslash}X>{\raggedright\arraybackslash}X}
\toprule
\textbf{Aspect} & \textbf{Level 1: push learning} & \textbf{Level 2: explicit feedback learning} & \textbf{Level 3: implicit feedback learning} \\
\midrule
\textbf{Feedback type} & None. Only curated content pushed into the agent. & Explicit. Natural language explanations logged as feedback on answers. & Implicit. Behavior traces and light signals such as clicks, dislikes, overrides, and heavy edits. \\
\addlinespace
\textbf{Memory structure} & Static corpora and configuration. Curated documents, templates, style guides, and tagged data. & Corpora plus feedback log per task or team. Each entry links an answer or fragment to a user explanation. & Corpora plus shared trajectory store (one common database where you keep the step-by-step histories of many agents run, and more than one agent can use those histories). Memory holds episodes: state, actions, outputs, user reactions, and rewards. \\
\addlinespace
\textbf{Learning logic} & Learning only through manual changes to content and prompts by designers. No automatic adaptation from usage. & Batch or incremental update from explicit corrections. Adaptation logic reads feedback log and adjusts prompts, retrieval rules, or small models. & Pattern discovery over episodes. Learning engine infers improvements from telemetry using heuristics, trial-and-error methods that test alternatives and shift toward whichever option performs best; learning methods that review full sequences of past actions and outcomes to improve future decisions. \\
\addlinespace
\textbf{Architecture impact} & Data and memory layer only. Interface stays simple, orchestration does not log feedback, and memory is read only during operation. & Data and interface layers both change. Interface collects structured feedback, memory stores it, and agent layer adds a small adaptation loop. & All three layers change. Interface emits implicit signals and telemetry, orchestration logs and tags each step, and data layer becomes a behavior and challenge store for agents. \\
\bottomrule
\end{tabularx}
\end{table}

As interviewees stressed, memory design---that is, access to static data and the history of interactions with the environment---must match learning maturity, because each step up in the autonomous learning level increases how the system adapts and how changes propagate, thereby raising governance requirements. For example, level 1 works best with simple, mostly read-only memory and light instrumentation. Level 2 expands memory to store structured, attributable user feedback, which requires interface features for capturing feedback and orchestration that links feedback to specific steps, tool calls, and sources. Level 3 treats memory as a comprehensive record of experience and implicit signals, which makes rich telemetry, step-level traces, strict access controls, versioning, and auditability essential---since the system may adapt without explicit human instruction and must remain traceable and accountable.

As variously emphasized by interviewees, data is the primary learning resource for agents, and data governance defines what an agent can know and what it is allowed to learn. The organization's agent knowledge base (the static data agents had access to) was deliberately bounded to trusted internal outputs, licensed external sources, and tightly controlled parts of the public internet. This key data governance decision was implemented because inconsistent access across tools produced inconsistent answers.

\subsection{Part C: Governance by Design for Agentic AI}

At the IT services firm, the gradual implementation of bounded assistants first, generative assistants after, and then agentic AI was driven by scaling challenges, which in turn were related to the learning maturity of individual agents and the system. Progress depended on whether the system could sustain measurable gains in cost, speed, research output quantity and quality, and risk mitigation, while resolving bottlenecks in continuity, integration, feedback, and control, which were highly dependent on memory and data configurations.

For example, early pilots could generate useful drafts or automate isolated steps, but scaling required predictable behavior, consistent access to evidence, and auditable execution across a growing portfolio of components. The organization encountered this constraint as a system-level coordination problem that cannot be treated as a separate policy layer added after prototyping; it must be implemented through governance mechanisms.

When agents can plan, call tools, and sequence multi-step work, architecture already embeds and enforces governance through access and rules: which sources are searchable, which tools are callable, what gets logged, how memory is stored, and where humans should intervene. The challenges we recount below show how these design choices became the practical mechanisms for achieving scalability, flexibility, and ongoing reliability, while also forcing explicit trade-offs among latency, accuracy, and operational costs. We highlight four such challenges---drawn from our case observations and interviews---and then discuss guided solutions implemented by the company.

\paragraph{Challenge 1: Continuity under missing memory.} The organization encountered ``missing memory'' as a governance problem when piloting multi-agent execution; delegated work became difficult to coordinate and verify because the system could not reliably carry state across steps, agents, and sessions. For example, if a user asked the system to analyze a dataset and then, a few hours later (or even less), asked it to refine the analysis, the system would often forget context from the first session. It couldn't recall which specific data tables it had used or the definitions it had assumed, because each agent run was isolated. This ``limited conversational memory across assets was a pain point'' (interviewee 4), the team noted after use. Researchers found themselves re-specifying information the system had output earlier, effectively having to remind the AI of its own recent work.

This challenge pushed the organization to confront a design question: How should memory be handled in a multi-agent, multi-tool environment? The team realized they needed some form of persistent context that outlived a single run. They began exploring solutions, debating between a centralized memory vs.\ a distributed memory approach. On the one hand, they considered having ``one common, monolithic memory that every asset talks to'' (a shared database in which all agents store and retrieve context). This could ensure continuity, so that any agent could see what any other had done. On the other hand, they saw appeal in ``per-agent memories that can securely read each other's context when needed'' (interviewee 11). This would isolate knowledge to each agent by default, but allow controlled sharing.

Each approach had drawbacks. A single memory could become both a governance headache (with questions like ``Who can edit it? What if sensitive info from one context bleeds into another?'') and a technical bottleneck (e.g., many reads/writes and potential for inconsistency). Multiple memories, on the other hand, would need protocols to synchronize or query one another, introducing coordination overhead and potential risks.

\medskip
\noindent$\rightarrow$ \textbf{Solution:} The IT services company deferred implementing full persistent memory, recognizing that the core design tension lay not in whether to have memory, but in how to distribute it: centrally (a single, shared store serving all agents) or in a distributed model (per-agent bounded memory). The centralized option maximizes coordination, but risks governance overload as cross-agent access rules become unmanageable at scale and can expose sensitive material across unrelated agents. The distributed option offers better containment, but agents can no longer passively draw on shared context, as they rely on information being explicitly passed between them, which adds coordination complexity. The team thus opted for an ``in between'' solution, which meant catching key data and IDs between steps and giving users the ability to re-upload or reference previous outputs if needed---while they continued to study the memory designs. The rationale was that the system (and users) were still at an early learning stage (level 1). As one AI engineer admitted, ``\emph{It's just very early stage, and the memory problem is a problem for later. Right now, we have very simple agents, so they don't need this big, shared memory thing}'' (interviewee 11). The company recognized that tackling full-blown, persistent memory would raise new issues (data-retention policies, privacy, cost) that they weren't ready to solve overnight.

\paragraph{Challenge 2: Spaghetti systems.} As the company scaled beyond bounded assistants, the fragmentation of tools, interfaces, and data silos became an empirical governance problem. When multiple interfaces and overlapping tools coexisted without a unified entry point, users could not easily determine which channel to use for a given task. This ambiguity increased the likelihood that governance controls would be bypassed, not deliberately, but because users defaulting to the most familiar or convenient tool would skip the channels where logging, review steps, and approval flows were built in---thus weakening accountability for both the evidence base and the execution trail.

The practical consequence was that work was split across fragmented dashboards and disconnected assistants, so neither the system nor the user could maintain a coherent state across runs. Users had to re-specify context at each entry point, meaning that no single record captured what had happened, who had acted, or on what basis, making it impossible to trace decisions end-to-end. The firm therefore came to see system design and organizational design as the same problem: Every architectural choice about where information lives, how outputs move between steps, and what gets logged was also a decision about how the system is embedded in the organization (i.e., which teams can access it, through which channels, for what purposes, and under what constraints).

\medskip
\noindent$\rightarrow$ \textbf{Solution: Centralize execution through a governed ``Nexus''.} The company's response was to consolidate interactions through a single governed entry point (an orchestrator agent, also called ``Nexus'') that routes requests behind the scenes and standardizes how access controls, data scoping (i.e., defining which data sources each agent is allowed to access for a given task), and tracking are applied, reducing governance bypass by making the controlled path the default path. Here, the organization should control not only what the model can generate, but also which tools and sources can be invoked and under what conditions. For instance, the organization introduced data agents that make sense of relevant data based on user prompts and an LLM.

In maturity terms, this solution aligns primarily with level 1 (push learning) because it establishes the prerequisite governance basis: a managed collection of trusted sources where the organization controls who owns each source, tracks changes over time, and sets access rights, and a controlled routing mechanism that enforces data governance at execution time. To progress to level 2, the organization would need structured feedback signals tied to specific routing and retrieval decisions (so the ``Nexus'' can learn under governance); to progress to level 3, the organization would need telemetry and trajectory data treated as governed datasets (retention, access, explainability), because centralization increases the volume and sensitivity of execution traces and behavioral signals.

\paragraph{Challenge 3: Learning from use without losing traceability and control over change.} As the company's agentic system scaled to more users, a new governance problem emerged: how to let the system improve, based on how real users were interacting with it once ``live'' (as opposed to the controlled conditions of design and testing) without losing control over what changes, when, and why. Usage logs revealed uneven adoption: Data-gathering tasks were widely adopted, while drafting functions were adopted more cautiously. Users frequently interrupted runs and reformulated their queries, signaling a mismatch between what the agent was doing and what the user needed. These patterns were informative, but they were not being systematically captured or fed back into the system in a structured way.

Governance risk emerged because the system's performance was evolving informally through user workarounds, rather than through structured learning mechanisms. The system lacked a structured way to remember corrections, preferences, or failure patterns across runs. Each interaction was largely isolated, which meant that learning remained manual and episodic rather than cumulative. Governance risk arose because behavioral corrections were not embedded into system memory in a traceable, auditable way.

Usage signals also revealed friction around flow, latency, and source quality. Some outputs were perceived as slow or poorly sequenced, and specific sources were flagged as outdated. Yet without structured feedback tied to retrieval traces or tool calls, engineers could not easily determine whether problems originated in data retrieval, prompt design, orchestration order, or model choice. This weakened traceability and limited the organization's ability to adjust data priorities or optimize tool routing in a defensible manner.

\medskip
\noindent$\rightarrow$ \textbf{Solution:} The company's workaround was to engineer feedback and intermediate visibility into the workflow, thereby capturing user ratings/comments, surfacing steps, and using corrections to drive controlled updates (e.g., prompt/library adjustments, tone settings, retrieval prioritization), rather than ad hoc tweaks. This solution constitutes the operational prerequisite for reaching level 2 (explicit feedback learning) in the company's roadmap and simultaneously defines what level 2 means.

Level 2 requires that the system receive structured signals from users, integrate them into its behavior in controlled ways, and do so without destabilizing the evidence base it relies on. None of that is possible without the infrastructure that the organization built: capturing ratings and corrections at the workflow level, surfacing intermediate steps so users can intervene meaningfully, and routing those signals into governed updates. This, in turn, implies a dual-memory design, where short-term working memory handles the live execution context of a given run, while a governed long-term feedback memory accumulates learning signals that are reviewed before integration. Without that separation, any adaptation would risk corrupting the stable knowledge base the system depends on, undermining the traceability the solution is meant to preserve.

\paragraph{Challenge 4: Balancing workflows and agents---cost, latency, and accountability trade-offs.} As the company's system matured, another governance tension emerged over when to use deterministic workflows and when to deploy autonomous agents. Workflows were ``predefined control flows'' that execute fixed sequences, suitable for routine, repeatable tasks such as standardized data updates. Agents, in contrast, could interpret goals, plan multi-step actions, dynamically select tools, and adapt as needed. This distinction was not merely technical but governance-relevant because autonomy directly affected cost, latency, traceability, and risk exposure.

Workflows required minimal memory since each run followed a stable script. Agents, especially in multi-agent settings, generated intermediate states and decisions that required memory to preserve coherence across steps. The more autonomy that was granted, the more sophisticated the memory needed to support planning and coordination, increasing governance complexity around how long data is stored, who can access it, and whether decisions can be traced back to their inputs.

Agentic reasoning also consumed more tokens (the basic units of text a language model processes and is billed for---roughly one per word). In other words, each planning step, tool invocation, and iterative refinement required additional API calls to the underlying language model, compounding computational cost and raising latency. Unconstrained autonomy risked unnecessary compute expenditure and slower performance, whereas workflows constrained tool paths and data use, offering predictability and efficiency. Governance, therefore, required explicit decisions about when flexible data retrieval and adaptive reasoning were justified.

\medskip
\noindent$\rightarrow$ \textbf{Solution:} Deterministic workflows were preserved for well-understood tasks where efficiency and auditability mattered most, while agents were reserved for complex, high-value problems requiring adaptive reasoning. This embeds cost control and accountability directly into the system architecture. The underlying principle is that autonomy should be conditional and aligned with governance maturity. Memory design was thus matched to ``execution'' mode: lightweight session memory for workflows and more advanced mechanisms for agentic components. In maturity terms, workflows align with lower learning levels and limited persistence, whereas agentic execution requires stronger logging, instrumentation, and feedback mechanisms before advancing to higher learning maturity. Autonomy expanded only where memory structures, cost controls, and oversight mechanisms were sufficient to govern it responsibly.

\bigskip
These four challenges and workarounds exposed a set of interconnected weaknesses in the initial setup: the absence of persistent memory, the fragmentation of data sources and tool interfaces, the lack of structured feedback mechanisms, and unclear criteria for deciding when a task should be handled by a deterministic workflow, versus an autonomous agent. Each weakness traced back to how data and memory were configured. When the system could not carry state across agents or sessions, governance rules requiring audit trails, verification, or continuity of evidence had nothing to which to anchor (challenge 1).

When data sources were scattered across disconnected tools, users bypassed the channels where logging, approval flows, and access controls were enforced (challenge 2). When the system had no structured way to capture user corrections and feed them back into decisions about which data sources to query, which tools to invoke for a given step, and how prompts should be adjusted, learning remained manual and ad hoc, and engineers could not pinpoint whether problems sat in retrieval, orchestration, or model output (challenge 3). And when the system lacked clear criteria for choosing between deterministic workflows and autonomous agents, it could not determine whether a given task should run as a fixed, predictable sequence or as a flexible, goal-directed agent with the governance infrastructure to support either option (challenge 4).

In short, data management is the ground on which every governance rule stands, since access controls define what agents can see, audit trails depend on the data they log, and output quality is bounded by the sources the system draws on. Memory configuration provides connectivity, giving the system a common space for information to be passed across sessions, agents, and tools, so that prior decisions can be recalled and corrections can accumulate. Meanwhile, learning maturity determines how safely the system can refine its own behavior as it encounters new tasks and unfamiliar situations.

These three elements form a reinforcing loop. The choice of memory architecture shapes which data the system can retain, which, in turn, determines what feedback signals are available for learning. As learning matures, it generates new demands on memory (longer retention, richer context, finer-grained traceability) and on data governance (broader access scope, stricter quality controls, clearer provenance). Conversely, when data is fragmented or ungoverned, memory has nothing reliable to store and resurface; when memory is limited, the system has nothing to learn from. Progress on any one element pulls the other two forward, while neglecting anyone creates a bottleneck that holds back the rest.

\citet{alaimo2024data} characterize such decisions as intrinsic to data's organizational role: By bounding the agent's epistemic horizon, organizations do not merely filter information but actively construct the conditions under which knowledge can be reliably produced and governed. The curation of data sources thus operates as an infrastructural practice \citep{bowker1999sorting} that shapes what the agent can know, learn from, and act upon. In complex and adaptive environments, curating and governing data \citep[see also][]{lycett2013datafication, alaimo2024data} therefore becomes a primary lever for controlling agent behavior, improving output quality, and managing compliance risk.

Taken together, the experience of the IT services organization demonstrates that scaling agentic AI requires sequencing these architectural decisions, so that governance capacity evolves alongside system capability. Throughout this progression, human authority shifts rather than disappears as the system becomes more capable; the human role moves from supervising every intermediate step to exercising control at decisive moments, such as defining task boundaries, approving execution plans, interpreting outputs, and deciding which corrections should feed back into the system. All together, these design choices transform agentic architecture into an organizational learning infrastructure in which the system gains greater independence only as its data, memory, and oversight foundations are ready to support it.

Below, we distill seven lessons for leaders seeking to implement and scale agentic AI systems.

\section{Lessons Learned}

The organization's transition to agentic AI offers rich lessons for any leader looking to implement similar systems. We distill seven key lessons, each with a brief explanation of why it matters, an example from the company's journey, and the decision that it points to.

\paragraph{Lesson 1: Adopt agentic AI with an ROI-first lens and treat scaling as organization-wide change.}

Agentic systems reshape how work is produced and coordinated across the firm. They change workflows, handoffs, team roles, verification routines, and which data sources become authoritative. As a result, adoption depends on change management as much as on engineering. Without a return-on-investment-first assessment, organizations risk adding another layer of tooling that increases routing confusion, duplicates existing technologies, and amplifies siloed or outdated data structures rather than resolving them (creating the ``spaghetti system'' described above).

The lesson learned from the case implies that agentic systems should be introduced only where they address a clear organizational bottleneck and where benefits can be measured---such as whether the system reduces the cost of producing a deliverable, shortens the time from request to finished output, allows the team to handle more projects with the same headcount, opens new revenue streams, or lowers the chance of errors and compliance failures. This evaluation should also anticipate how work will be reorganized: which tasks will follow fixed, predictable sequences and which will be handled by autonomous agents; how responsibilities and approval flows will change; and whether the system will bring scattered data sources together under a single, governed entry point or deepen existing fragmentation.

\emph{Decision:} Begin with a concrete ROI hypothesis tied to a specific workflow problem, along with measurements of current performance before the system is introduced. Compare the agentic approach to existing tools and processes, define the target operating model for teams and data, and prioritize resource change management explicitly through training, local champions, revised workflows, and governance routines that support adoption and controlled scaling.

\paragraph{Lesson 2: Start with controlled co-learning before chasing full autonomy.}

Jumping directly to highly autonomous agents can overwhelm organizations and make governance and accountability hard to see, especially when systems can plan, call tools, and change behavior over time. The lesson learned from the case recommends treating implementation as a staged, co-learning journey for both users and the system, where autonomy expands only as controls and practices mature. The practical implication is to roadmap capability in levels and tie each step to concrete guardrails, such as the set of sources the system is permitted to draw on when producing an output, memory rules, instrumentation, and approval gates, so that trust and accountability scale with capability.

This staged approach also supports rapid evolution. Today, models, tools, and agent capabilities are changing faster than traditional enterprise technology cycles, so agentic systems should be built with flexibility in mind from the outset. In practice, that means modular components, replaceable tools, versioned prompts and workflows, and governance routines that allow frequent iteration without uncontrolled drift.

\emph{Decision:} Plan Agent and AI capability in stages, monitoring and adjusting memory and data needs along the way. Start with your AI as a junior partner (e.g., level 1 maturity) that is fully under human oversight and only grant it more autonomy as your team builds trust and you have put new guardrails in place. Agentic systems should be built with flexibility at their core, as models, tools, and agent capabilities evolve faster than traditional enterprise technology cycles.

\paragraph{Lesson 3: Use your company data as both the training material and the boundary for what the AI can rely on.}

The AI's output quality and compliance are determined by the data it is allowed to use. Leaders often focus on the model, but the case shows that the crux is a well-governed data supply. When data sources were scattered across disconnected tools and teams, the same question could produce different answers depending on which tool a user asked, because each drew on a different slice of the organization's knowledge. Centralizing data sources and eliminating data silos improved consistency and enabled enforcement of a single standard for what counts as a trusted source. The company also used filters so that the AI only retrieved from licensed or vetted content, drastically reducing misinformation and legal risk.

This design choice reflects a deeper insight: Data are not neutral raw materials, but rule-governed objects whose meaning and usefulness depend on how they are classified, standardized, and maintained \citep{alaimo2024data}. By restricting agent access to vetted, internally curated sources, the organization defines what counts as valid evidence and how it may be combined, transformed, and reused. Data governance in agentic systems also has a self-reinforcing quality, since the data agents retrieve and shape their outputs, which, in turn, become inputs for subsequent agent actions and organizational decisions. Governing data is therefore not a one-time policy decision, but an ongoing organizational practice that is embedded in the architecture itself.

\emph{Decision:} Invest in your data foundation before algorithms. Identify the sources your AI will draw from, clean and tag them, and enforce permissions. Think of it as creating the AI's library and rulebook, as it can't read (or misread) what's not in the library. Ensure data governance (i.e., licensing, privacy, quality) is built into AI design decisions from the start, and consider a cross-functional data management team to decide what your AI can see or ignore.

\paragraph{Lesson 4: Embed oversight into the workflow.}

If you only check AI outputs at the end, it's often too late or too costly to fix issues. The company's approach shows the value of real-time governance: putting human approval steps and automated checks at critical points, so that errors and misdirection are caught early. For example, in the company's system, no complex task runs without the user first reviewing the AI's plan and source scope. The UI literally requires a click to proceed, effectively turning policy into practice. Similarly, AI reviewer agents continuously monitor outputs for factual accuracy, serving as built-in auditors.

For leaders, this means working with your product teams to include features like approval gates for high-impact actions, explainability panels that show how the AI arrived at a result, and logs that are easy to audit. Scaling agentic systems requires a stronger ``human-in-the-lead'' stance, where people remain accountable by governing the system's goals and boundaries through design choices that determine what tasks are permissible, how much autonomy is granted, which workflows are allowed, what data and tools the system can access, what is logged and retained, and when changes to behavior are introduced.

\paragraph{Lesson 5: Make system memory a deliberate choice, aligned with accountability.}

What the AI remembers (or forgets) determines how errors propagate and how learning accumulates. Short-term memory systems may cause the AI to forget or become repetitive; long-term, as in a monolithic stack, might be too costly and resource-intensive, and could run away with unchecked learning or expose sensitive information. For example, they decided that at the AI's current maturity, it would not retain long-term memories across sessions unless explicitly allowed. This prevented compounding errors and kept each session accountable to its inputs. In contrast, as the company plans for more advanced learning, they know they'll need to introduce shared memory or logs, but with audit trails.

This is both a design and policy matter. For instance, it is advisable to set retention policies for AI-generated content and conversations, similar to those for human knowledge management. If persistent learning is enabled (e.g., model fine-tuning or adaptive prompts based on history), it is also wise to enable a way to review and roll back those learnings if needed. Leaders should insist on transparency, asking their developer teams: ``What is our AI storing from user interactions? And how do we inspect or erase it if required?''

\paragraph{Lesson 6: Operational cost awareness coupled with latency vs.\ accuracy.}

Agentic systems create recurring costs in computing, orchestration, monitoring, and storage. Token spend increases when agents plan, iterate, call tools, and carry long context---and telemetry and audit logging add ongoing costs for data pipelines and retention. If these costs are not built into the business case early, scaling can become financially and operationally brittle. Our case shows that workflow versus agentic execution is a cost-and-control decision. Deterministic workflows fit fast and repeatable tasks because they reduce token-intensive reasoning, improve latency, and remain easier to test and audit, but may be less accurate. Agentic execution should be reserved for tasks that require adaptive planning and multi-step coordination, where the added cost produces measurable value.

\emph{Decision:} Organizations should define decision points at which leaders determine whether speed or precision takes priority for a given task, as well as apply higher-accuracy settings and deeper verification only to high-impact or high-risk tasks. Set task cost and latency budgets, track token spending by workflow and agent, and specify which tasks run as deterministic workflows versus agentic execution. Additionally, organizations should manage context length, memory persistence, and telemetry depth as cost levers, because the amount of prior conversation and data that the system carries into each step improves coherence in longer contexts, but increases the number of tokens processed, thereby raising computing costs.

\paragraph{Lesson 7: Scalability as an architectural and organizational constraint.}

Agentic pilots often succeed in controlled environments, yet scaling them across functions, geographies, and volumes introduces new architectural and organizational tensions. At the company, scalability was constrained not only by technical complexity but also by organizational bandwidth. For example, the agentic lab team was frequently pulled into client-facing engagements, which disrupted asset design-and-build workflows and slowed systematic stabilization.

A further tension emerged between deterministic workflows and sophisticated agents. General agents are easier to scale across teams and contexts because they rely on reusable logic and standardized access to tools. In contrast, agents adapted to specific processes may deliver higher local performance but are harder to maintain, replicate, and govern across organizational units. Without deliberate architectural boundaries, scaling can lead to fragmentation, duplicated effort, and inconsistent governance practices.

\emph{Decision:} Organizations should treat scalability as a design constraint from the outset, rather than an afterthought. This means defining whether an agent is intended to be generalizable infrastructure or context-specific augmentation, and then aligning memory design, tool routing, and data boundaries accordingly. It's also advisable to establish clear ownership and capacity allocation for core agentic assets to prevent delivery pressures from destabilizing foundational design work. Ditto for standardizing reusable components, interfaces, and governance controls where cross-functional scale is intended.

\bibliographystyle{plainnat}

\begin{thebibliography}{19}
\providecommand{\natexlab}[1]{#1}
\providecommand{\url}[1]{\texttt{#1}}
\expandafter\ifx\csname urlstyle\endcsname\relax
  \providecommand{\doi}[1]{doi: #1}\else
  \providecommand{\doi}{doi: \begingroup \urlstyle{rm}\Url}\fi

\bibitem[Alaimo and Kallinikos(2024)]{alaimo2024data}
Cristina Alaimo and Jannis Kallinikos.
\newblock \emph{Data Rules: Reinventing the Market Economy}.
\newblock MIT Press, 2024.
\newblock \doi{10.7551/mitpress/11751.001.0001}.

\bibitem[Ananny and Crawford(2018)]{ananny2018seeing}
Mike Ananny and Kate Crawford.
\newblock Seeing without knowing: Limitations of the transparency ideal and its
  application to algorithmic accountability.
\newblock \emph{New Media \& Society}, 20\penalty0 (3):\penalty0 973--989,
  2018.

\bibitem[Berente et~al.(2021)Berente, Gu, Recker, and
  Santhanam]{berente2021managing}
Nicholas Berente, Bin Gu, Jan Recker, and Radhika Santhanam.
\newblock Managing artificial intelligence.
\newblock \emph{MIS Quarterly}, 45\penalty0 (3):\penalty0 1433--1450, 2021.

\bibitem[Bowker and Star(1999)]{bowker1999sorting}
Geoffrey~C. Bowker and Susan~Leigh Star.
\newblock \emph{Sorting Things Out: Classification and Its Consequences}.
\newblock MIT Press, 1999.

\bibitem[Dignum(2019)]{dignum2019responsible}
Virginia Dignum.
\newblock \emph{Responsible Artificial Intelligence: How to Develop and Use AI
  in a Responsible Way}.
\newblock Springer, 2019.

\bibitem[Floridi et~al.(2018)Floridi, Cowls, Beltrametti,
  et~al.]{floridi2018ai4people}
Luciano Floridi, Josh Cowls, Monica Beltrametti, et~al.
\newblock {AI4People}---An ethical framework for a good {AI} society.
\newblock \emph{Minds and Machines}, 28\penalty0 (4):\penalty0 689--707, 2018.

\bibitem[{European Parliament and Council}(2024)]{euaiact2024}
{European Parliament and Council}.
\newblock Regulation ({EU}) 2024/1689 laying down harmonised rules on
  artificial intelligence ({AI Act}), 2024.

\bibitem[Lewis et~al.(2020)Lewis, Perez, Piktus, et~al.]{lewis2020retrieval}
Patrick Lewis, Ethan Perez, Aleksandra Piktus, et~al.
\newblock Retrieval-augmented generation for knowledge-intensive {NLP} tasks.
\newblock \emph{Advances in Neural Information Processing Systems}, 33, 2020.

\bibitem[Lycett(2013)]{lycett2013datafication}
Mark Lycett.
\newblock `{D}atafication': Making sense of (big) data in a complex world.
\newblock \emph{European Journal of Information Systems}, 22\penalty0
  (4):\penalty0 381--386, 2013.

\bibitem[Mikalef and Gupta(2021)]{mikalef2021artificial}
Patrick Mikalef and Manjul Gupta.
\newblock Artificial intelligence capability: Conceptualization, measurement
  calibration, and empirical study on its impact on organizational creativity
  and firm performance.
\newblock \emph{Information \& Management}, 58\penalty0 (3):\penalty0 103434,
  2021.

\bibitem[{NIST}(2023)]{nist2023ai}
{NIST}.
\newblock {AI} Risk Management Framework ({AI RMF} 1.0).
\newblock Technical report, National Institute of Standards and Technology,
  2023.

\bibitem[Orlikowski(2007)]{orlikowski2007sociomaterial}
Wanda~J. Orlikowski.
\newblock Sociomaterial practices: Exploring technology at work.
\newblock \emph{Organization Studies}, 28\penalty0 (9):\penalty0 1435--1448,
  2007.

\bibitem[Rahwan(2018)]{rahwan2018society}
Iyad Rahwan.
\newblock Society-in-the-loop: Programming the algorithmic social contract.
\newblock \emph{Ethics and Information Technology}, 20\penalty0 (1):\penalty0
  5--14, 2018.

\bibitem[Raisch and Krakowski(2021)]{raisch2021artificial}
Sebastian Raisch and Sebastian Krakowski.
\newblock Artificial intelligence and management: The automation--augmentation
  paradox.
\newblock \emph{Academy of Management Review}, 46\penalty0 (1):\penalty0
  192--210, 2021.

\bibitem[Raji et~al.(2020)Raji, Smart, White, et~al.]{raji2020closing}
Inioluwa~Deborah Raji, Andrew Smart, Rebecca~N. White, et~al.
\newblock Closing the {AI} accountability gap: Defining an end-to-end framework
  for internal algorithmic auditing.
\newblock In \emph{Proceedings of the 2020 Conference on Fairness,
  Accountability, and Transparency (FAccT)}, 2020.

\bibitem[Schick et~al.(2023)Schick, Dwivedi-Yu, Dess{\`i},
  et~al.]{schick2023toolformer}
Timo Schick, Jane Dwivedi-Yu, Roberto Dess{\`i}, et~al.
\newblock Toolformer: Language models can teach themselves to use tools.
\newblock \emph{arXiv preprint arXiv:2302.04761}, 2023.

\bibitem[Selbst et~al.(2019)Selbst, Boyd, Friedler, Venkatasubramanian, and
  Vertesi]{selbst2019fairness}
Andrew~D. Selbst, Danah Boyd, Sorelle~A. Friedler, Suresh Venkatasubramanian,
  and Janet Vertesi.
\newblock Fairness and abstraction in sociotechnical systems.
\newblock In \emph{Proceedings of FAT*}, 2019.

\bibitem[Vassilakopoulou et~al.(2022)Vassilakopoulou, Hustad, and
  Olsen]{vassilakopoulou2022information}
Polyxeni Vassilakopoulou, Eli Hustad, and Dag~H{\aa}kon Olsen.
\newblock Information systems research on artificial intelligence: A call for
  sociotechnical perspectives.
\newblock \emph{Journal of the Association for Information Systems},
  23\penalty0 (2):\penalty0 506--531, 2022.

\bibitem[Wang et~al.(2024)Wang, Ma, Feng, Zhang, Yang, Zhang, Zhao, Wei, Wen,
  et~al.]{wang2024survey}
Lei Wang, Chen Ma, Xueyang Feng, Zeyu Zhang, Hao Yang, Jingsen Zhang, Wayne~Xin
  Zhao, Zhewei Wei, Ji-Rong Wen, et~al.
\newblock A survey on large language model based autonomous agents.
\newblock \emph{Frontiers of Computer Science}, 18:\penalty0 186345, 2024.
\newblock \doi{10.1007/s11704-024-40231-1}.

\end{thebibliography}

\end{document}